\documentclass[prb,twocolumn,aps,showpacs,amsmath,amssymb,floatfix]{revtex4}

\usepackage{graphicx}

\begin{document}


\title{Freezing  effects in the  two dimensional  one-component plasma
 and in thin film type II superconductors}

\author{P. A. McClarty and M. A. Moore}
\affiliation{School of Physics and Astronomy,
 University of Manchester, Manchester M13 9PL, U.K.}  

\date{\today}

\begin{abstract}
We present results  of Monte Carlo simulations of  the two dimensional
one-component plasma  and of the  Ginzberg-Landau model in  the lowest
Landau   level  approximation,   with  both   charges   and  vortices
respectively confined  within a disc.  In both  models we see that
as  the temperature  is reduced,  oscillations in  the  radial density
develop which  spread into  the bulk  from the edge  of the  disc. The
amplitude of  these oscillations grows  as the temperature  is lowered
and the length scale over which the oscillations occurs is the same as
the   correlation  length   for  local   crystalline  order   at  that
temperature.  At temperatures  similar to those  where earlier studies have
reported a first-order fluid-crystal phase transition, the correlation
length is comparable to the  linear dimensions of the samples studied,
which suggests that finite size effects will be affecting the accuracy
of their conclusions.
\end{abstract}


\maketitle


In this  paper, we report  on freezing-related effects in  two models
with long-ranged interactions between  \lq particles', both of which have
some bearing on the  behavior of clean thin-film superconductors. The
models  are,  the  one-component  plasma  in two  dimensions  and  the
Ginzberg-Landau  model  in  the  lowest  Landau  level  approximation;
henceforth 2D OCP and the GL LLL models respectively.

Published simulations  of these models have  given conflicting results
regarding even the existence  of the fluid-crystal transition. For the
2D   OCP  with   periodic  boundary   conditions,  Leeuw   and  Perram
\cite{Leeuw}  found a  first  order transition  at about  $\Gamma=140$
(with that parameter  defined below) which showed up  as hysteresis in
the  energy when  the system  is cooled  and then  heated  through the
transition temperature.   The same conclusion was  reached by Choquard
and Clerouin \cite{Choquard} in  simulations of the disc geometry with
some particles close to the boundary fixed at vertices of a triangular
lattice.  In  contrast,  Monte  Carlo  simulations for  charges  on  a
spherical surface by Moore and P\'{e}rez-Garrido \cite{Moore} found no
evidence  for  a  transition;   instead  the  correlation  length  for
short-range  crystalline order  in  the  fluid was  found  to grow  to
infinity  as  the  temperature   was  reduced  to  zero.  A  dynamical
investigation of  the disc geometry, this time  without any constraint
on  the particle  positions has  been made  by Ying-Ju  Lai and  Lin I
\cite{Lai}.   They  looked at  the  dislocation  density in  different
regions  and suggested  that  the  system melts  from  the outside  as
dislocations proliferate around  the intrinsic disclinations which are
close to the boundary at low temperatures. They found that the central
region  loses  translational  and  then  orientational  order  as  the
temperature increases.

For  the GL  LLL model,  Monte Carlo  simulations  for two-dimensional
 periodic  cells  \cite{Xing,Kato,Hu}  have  supplied evidence  for  a
 first-order  fluid-crystal   transition  at  an   effective  \lq \lq
 temperature''  $\alpha_{T}=-10$   ($\alpha_{T}$  is  defined  below).
 Simulations on  a spherical surface \cite{Dodgson} and  within a disc
 \cite{ONeill}  find only  a fluid  phase with  a  growing correlation
 length for decreasing effective temperature.

Because  whether  one sees  a  phase transition  or  not  seems to  be
dependent  upon the boundary  conditions used,  we have  revisited the
problem to  investigate how finite  size effects may be  affecting the
outcomes of these diverse  simulations. In particular, we have studied
the  radial density for  the 2D  OCP and  the GL  LLL models  when the
particles  are confined  to a  disc. We  have found  that  it develops
oscillations or ripples of  increasing amplitude as the temperature is
reduced which reach in from the edge  of the disc by a distance of the
order  of the  length scale  of short-range  crystalline order  in the
system at that temperature, as found  for example in the work of Moore
and   P\'{e}rez-Garrido   \cite{Moore}.    The   existence   of   such
oscillations has been found in  other fluid systems in the presence of
an interface \cite{Evans, Evans2}.

Perhaps  the most  straightforward way  of seeing  the origin  of the
oscillations induced  in the density by  the presence of  a surface is
via linear response theory. A small external potential $\delta
\phi(\mathbf{r})$,     whose    Fourier    transform     is    $\delta
\hat{\phi}(\mathbf{k})$,  will  induce a  change  in the  one-particle
density $\delta \rho(\mathbf{r})$. Its Fourier transform is given by
 \cite{Hansen}
\begin{equation}
 \delta  \hat{\rho}(\mathbf{k})   =  -\beta\rho  S(\mathbf{k})  \delta
 \hat{\phi}(\mathbf{k}) \
\label{LR}
\end{equation}
so  the density  changes $\delta \rho(\mathbf{r})$  induced by  the potential
will  depend  on  the  analytic  structure in the
complex $\mathbf{k}$ plane of  the  structure  factor
$S(\mathbf{k})\equiv\langle
\rho(\mathbf{k})\rho(\mathbf{-k})\rangle/N$ of the  bulk system.
 It is argued that at distances well away from 
the surface its effects can be treated as if it corresponded to a perturbation applied at the edge of the system. 
When there is local crystalline order, the
structure  factor has  a peak  close to  the magnitude of the first  reciprocal lattice
vector, say at $|\mathbf{K}_{0}|$ and the profile of the peak is roughly
Lorentzian
\[ S(k) \sim \frac{1}{\left( k - K_{0} \right)^{2} + 1/\xi^{2} }, \]
as seen in earlier work \cite{Moore}, 
with $\xi$ the bulk correlation length, which has poles at $k = K_{0}\pm i/\xi$.
 The imaginary part controls the length scale of decay of the density oscillations. 
At large distances $d$ from the edge of the disc
\begin{equation}
 \delta \rho (d) \sim \exp(-d/\xi)\cos(K_{0} d). 
\label{oscillations}
\end{equation}

We first introduce the models, and then present our results on the
 surface induced  effects in a disc geometry. Finally we mention the
implications of these results for numerical simulations of the 2D OCP and the GL LLL model around the reported phase transition.


The 2D OCP  is a system of particles  interacting logarithmically in a
central  harmonic  confining  potential.  We arrive  at  the  harmonic
confining potential naturally by  finding the energy of interaction of
each charge with a background  charge density which is spread out over
the whole plane. Suppose we have  $N$ identical charges $e$ on a fixed
background of uniform charge density which confines them within a disc
of radius  $R$. It can be  shown that the
energy $V$ of  a configuration of $N$ charges  each labeled by integer
$i$ and with position $\left\{\mathbf{r}_{i}\right\}$ is, up to constants,
\begin{multline} \frac{V}{kT} = -\Gamma\sum_{i<j}^{N} \ln \left(
\left|\mathbf{r}_{i} - \mathbf{r}_{j} \right| \right) + \frac{\Gamma
  N}{2}\sum_{i=1}^{N}\left( \frac{\left| \mathbf{r}_{i} \right|
 }{R} \right)^{2}. 
\end{multline}
The temperature parameter that we use is $\Gamma = e^{2}/kT $. It is
  convenient to set the average number density $N/\pi R^{2}$ to $1/\pi$. In the London regime of thin film type II
  superconductors, which is characterized by low fields and hence low vortex densities so that variations in the complex order
  parameter can be neglected, vortices also interact logarithmically as in the 2D OCP.

The Ginzberg-Landau free energy functional for superconductors at
temperature $T$ depends on a spatially varying, complex order parameter
$\psi\left( \mathbf{r} \right)$, which is coupled to a magnetic vector potential $\mathbf{A}$. The
superconductor is exposed to a uniform and static magnetic field $B$ in the $\zeta$ direction. This field is close to the upper critical field
(deduced from the mean field approximation) so that it is almost uniform within the sample; we neglect the constant free magnetic
field energy term.    
\begin{multline*} F = \int d^{3}\mathbf{r} \left\{
  \frac{\hbar^{2}}{2m} \left| \left(
  -i\hbar\mathbf{\nabla} - e^{\star}\mathbf{A} \right)\psi
  \right| ^{2} \right. \\ \left. + \alpha\mid\psi\mid^{2} +
  \frac{\beta}{2}\mid\psi\mid^{4} \right\}.
\end{multline*}
The integral is taken over the entire sample volume. The parameters
$\alpha$, $\beta$, and $m$ are related to the temperature dependent
penetration depth and correlation length of the superconductor. We
shall be interested in a sample that is sufficiently thin (thickness $t$) in the
direction of the field that we can neglect any variation in the order
parameter in this direction and remove the integral over $\zeta$. Then
we restrict the order parameter to a linear combination of Landau
levels of lowest energy. In the symmetric gauge, with complex coordinates ($A_{x} = - By/2$ and $A_{y}= Bx/2$),  
 these take the form
\[ \phi_{l}\left( z\right) = g_{l}z^{l}e^{-\left| z \right|^{2}/4P}, \]
where $ l = 0,1,\dots $ and $ g_{l} = \left(\pi
 m!\right)^{-1/2}\left(1/2P\right)^{(m+1)/2} $ are normalization coefficients. $P=\hbar/ B$ is the squared magnetic length. The order parameter is 
\[ \psi\left( z \right) = Q \sum_{l=0}^{N}v_{l}\phi_{l}\left( z
\right), \]
where $Q=(2\pi P/\beta t)^{1/4}$ is introduced now to obtain a
 neater expression later.
The sum runs over the $N+1$ lowest angular momentum states. This is a
polynomial of degree $N$ in $z$ which therefore has $N$ zeroes
corresponding to the vortex positions in our finite system. This is
substituted into the free energy functional.  When the integrals have
been performed we get      
\begin{multline*} F = \alpha_{T}\sum_{m=0}^{N} v_{m}v_{m}^{\star} \\ + \sum_{p,q,r,s =
  0}^{N} 2^{m+p+2}\frac{\left( p+q \right)!}{\left( p!q!r!s!
  \right)}^{\frac{1}{2}}v_{m}v_{p}v_{n}^{\star}v_{r}^{\star}\delta_{m+p,n+r},
\end{multline*}
where $\alpha_{T}= Q^{2}t(\alpha+(e^{\star}B\hbar/m))$ \cite{Ruggeri, ONeill}.
Within  this approximation, the vortex matter has identical properties
along each of a family of curves labeled by $\alpha_{T}$ in the
magnetic field - temperature plane.
The quantity $\exp\left( -F \right)$ is taken as a Boltzmann weight in
the calculation of the canonical partition function $Z$. Several
authors \cite{Eilenberger,Urbach} have shown that at sufficiently high field in Type II superconductors,
fluctuations in the magnetic field can be neglected.

As   for   our  simulations,   we   used   the  Metropolis   algorithm
\cite{Metropolis}   to   approximate   thermal  averages   by   making
pseudorandom steps in the  coefficients $\left\{ v_{i}\right\}$ in the
Ginzberg-Landau simulations or in the  positions of the charges in the
OCP  simulations. We  thermalized the  sample  for a  period and  then
started  taking measurements  of configurations  to contribute  to the
averages.

\begin{figure}
\includegraphics[width=0.9\columnwidth,clip]{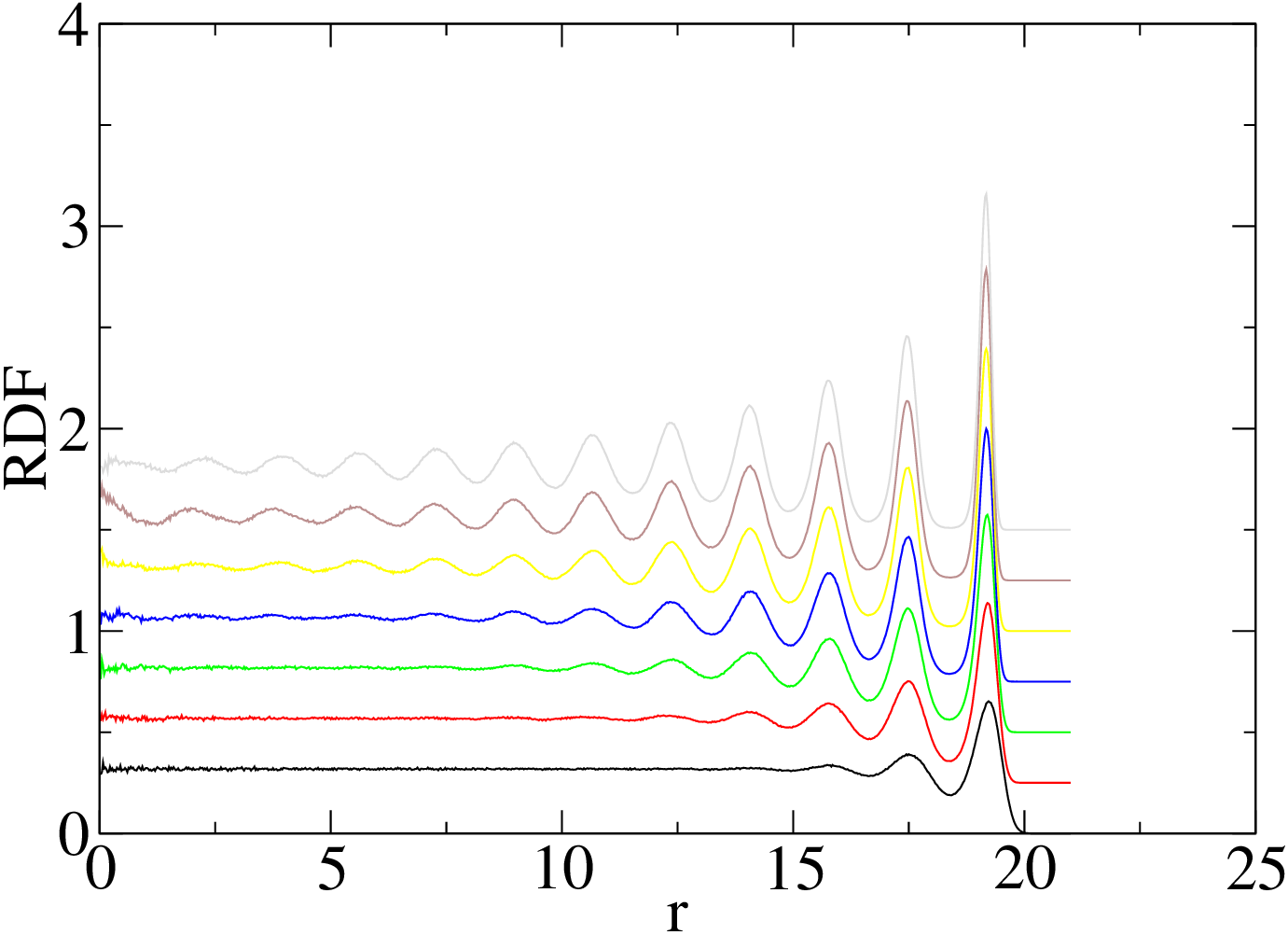}
\caption[Radial particle densities for the 2D one-component plasma at various temperatures and 400 charges.]{Radial particle densities for the 2D one-component plasma at
  $\Gamma=20$, $40$, $60$, $80$, $100$, $120$ and $140$ for $400$ charges
  showing progressive ordering from the edges. We have separated the curves along the vertical axis for the sake of clarity.}
\label{fig:OCPRD} 
\end{figure}

We computed the radial density for the 2D OCP as the thermal average of a fine-grained histogram of
particle numbers in circular shells for different discrete values of
$r$. As is known from the exact solution for $\Gamma = 2$, the radial
density is uniform from the center to the edge at a value of $1/\pi$. At lower temperatures,
oscillating behavior creeps in from the edge of the sample.  With
$N=400$ the pattern spreads across the system for $\Gamma=100$ though the
amplitude of the peaks continues to rise as the temperature is
lowered (Fig. \ref{fig:OCPRD}). We find, on increasing the number  of steps in  the simulation that these features do not change,
indicating that our system has been adequately equilibrated for their study.


We first consider the outermost peak. For $\Gamma$  greater than
about $100$ this peak corresponds well  to  an isolated
shell (with  a fixed  number of particles).  This  is because the density on either  side of the outer
peak  drops almost  to zero,  so there  must be  very  little particle
exchange between this  peak and neighboring ones. We  have found that
the height of this peak is proportional to $\sqrt{\Gamma}$. The reason
for this is  as follows. The radial oscillations  of particles in this
shell have an  amplitude $X$ that is equal to half the  width of the outer
shell. The  average energy of  this mode $e^{2}\langle  X^{2} \rangle$
is, by equipartition,  of order $kT$. So the  mean displacement of the
particles    from    equilibrium    is   roughly    proportional    to
$1/\sqrt{\Gamma}$ and hence, the height of the peak is proportional to
$\sqrt{\Gamma}$. The other peaks do  not obey this relation as closely
as  the outermost  peak because  the rest  of the  sample  has roughly
triangular order and so the  assumption of isolated shells with radial
oscillations breaks down.

\begin{figure}
\includegraphics[width=0.9\columnwidth,clip]{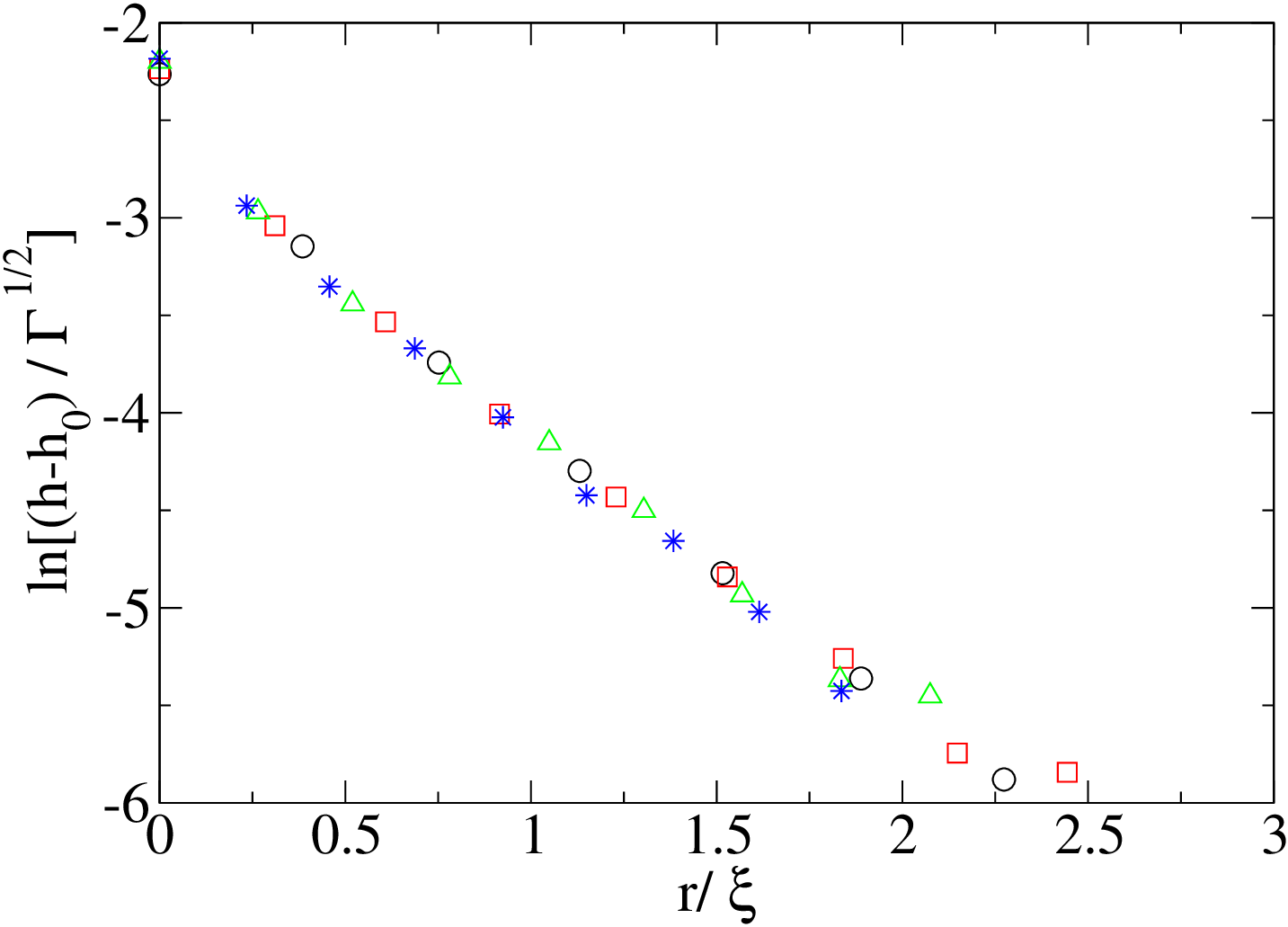}
\caption{Plot of $(h-h_{0})/\sqrt{\Gamma}$ against $r/\xi$ where $h$ are peak heights in the radial density, $a$ is the mean value of the radial
density and $r$ is the radial distance from the edge of the disc (as measured from the maximum peak). The correlation length $\xi =
\sqrt{\Gamma}-\beta$ with fitting parameter $\beta=4.46$. Plot includes data for $\Gamma=80$, $100$, $120$ and $140$.}
\label{fig:scaling} 
\end{figure}

It is  natural to  expect that  the peak heights  in the
radial density will decay with distance from the edge of the 
disc on a length scale which  is a reflection of the short-range
crystalline order in the system and that the length scale involved
is  just  the  correlation length  $\xi$  of  the  2D OCP.  Moore  and
P\'{e}rez-Garrido \cite{Moore} have  studied the structure factor
 $S(\mathbf{k})$ of
the   2D  OCP   on   the  surface   of   a  sphere   and  found  
 that from the inverse width of its first peak that the correlation length is
$\xi\sim\sqrt{\Gamma}$ at  low temperatures: that is, when  $\Gamma >
60$. We make the assumption that the same behavior for $\xi$ occurs in our disc
geometry.  $h-h_{0}$  is the  radial density  peak  height above  the
uniform  radial density $h_{0}$ which is $\approx 1/\pi$. We plot
  $\ln[(h-h_{0})/\sqrt{\Gamma}]$ against
$r/\xi$ where $r$ is the distance from the outer peak where the $\sqrt{\Gamma}$ scaling is based on
 the variation of the absolute height of
the outermost peak. The correlation length is  taken to vary  as $\sqrt{\Gamma}-\beta$ where $\beta$  is a
fitting parameter. Fig. \ref{fig:scaling}  shows collapse of data at
various  temperatures onto  a straight  line revealing  that  the peak
heights vary  exponentially with distance  from the highest  peak with a 
characteristic length that varies with  temperature in the same way as
on a spherical surface for $\Gamma\geq 60$. The parameter $\beta$ that
gives the best collapse is also consistent with already published data
for charges on a spherical surface \cite{Moore}.  

Because  the  rather general  considerations  of  Evans and  coworkers
 \cite{Evans}   give   oscillating    exponential   behavior   as   in
 Eq. \ref{oscillations}  only asymptotically  at large $d$,  we should
 not be surprised  that at least the first  peak height is exceptional
 in Fig. \ref{fig:scaling}, nor that  the peaks close to the edge have
 varying width and shape.  Indeed, if  this relation were  to hold close  to the
 edges with $\sqrt{\Gamma}$ scaling  of the heights, the density would
 not be completely positive even at quite high temperatures. We expect
 the  oscillating  exponential to  be  a good  fit  to  our data  only
 asymptotically.

\begin{figure}
\includegraphics[width=0.9\columnwidth,clip]{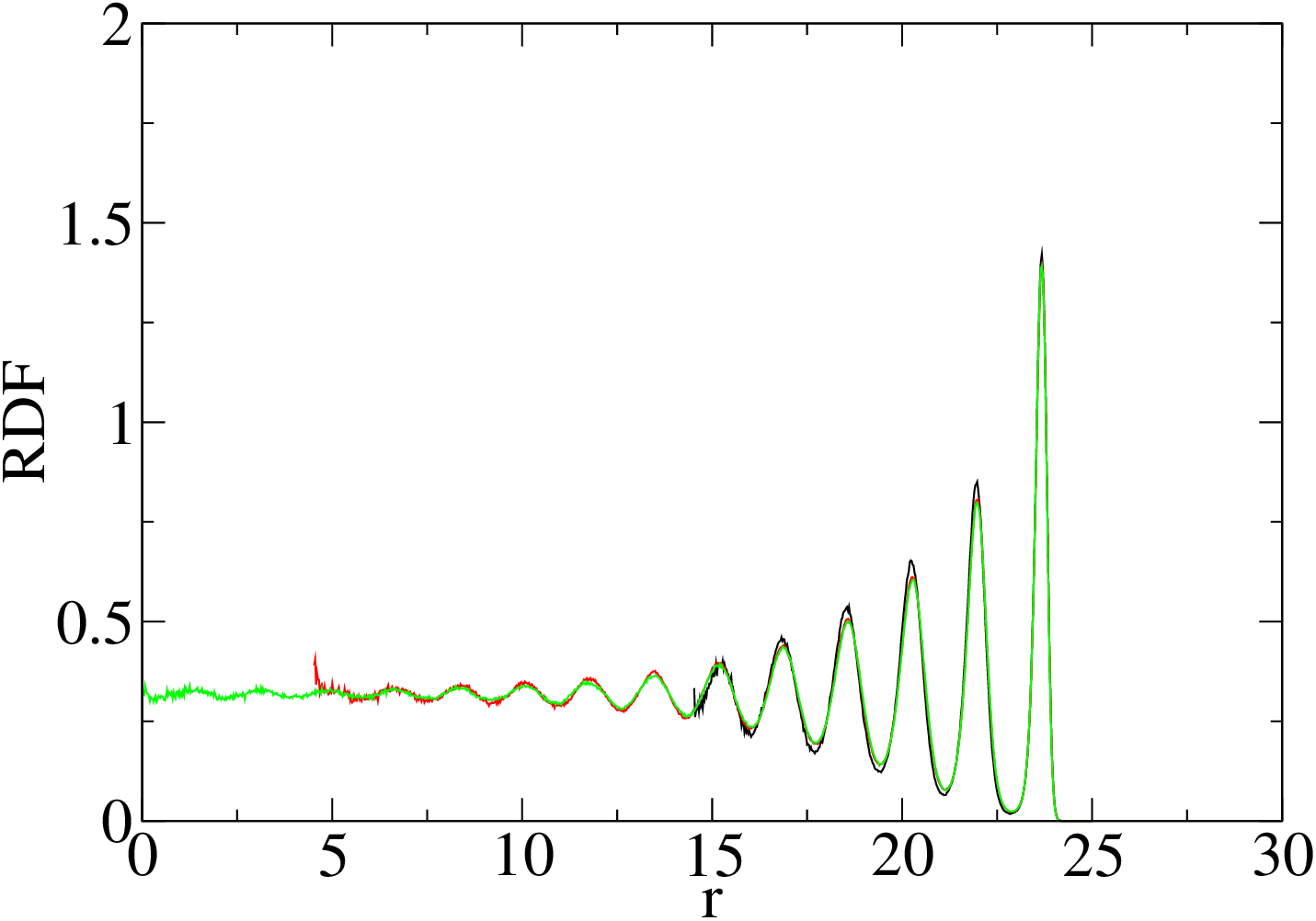}
\caption[Superimposed OCP radial densities for different system sizes.]{Two plots of radial densities for different system sizes ($100$, $400$, and $600$ charges)
  superimposed with the edge peak superimposed. The plot shown is for $\Gamma=100$.}
\label{fig:OCPFS} 
\end{figure}

The  data  in  Fig. \ref{fig:scaling}  were  taken
  for  a disc  containing  $400$
charges.  We  have superimposed 2D OCP  radial  densities for  identical
temperatures  and different  system sizes  with  the outer  peak as  a
reference  point  (Fig.  \ref{fig:OCPFS}).  The  superimposed  plots
overlap almost perfectly  independent
of the system size, indicating that the oscillations themselves are not finite size effects,
but are instead related to the presence of an edge in the system.


\begin{figure}
\includegraphics[width=0.9\columnwidth,clip]{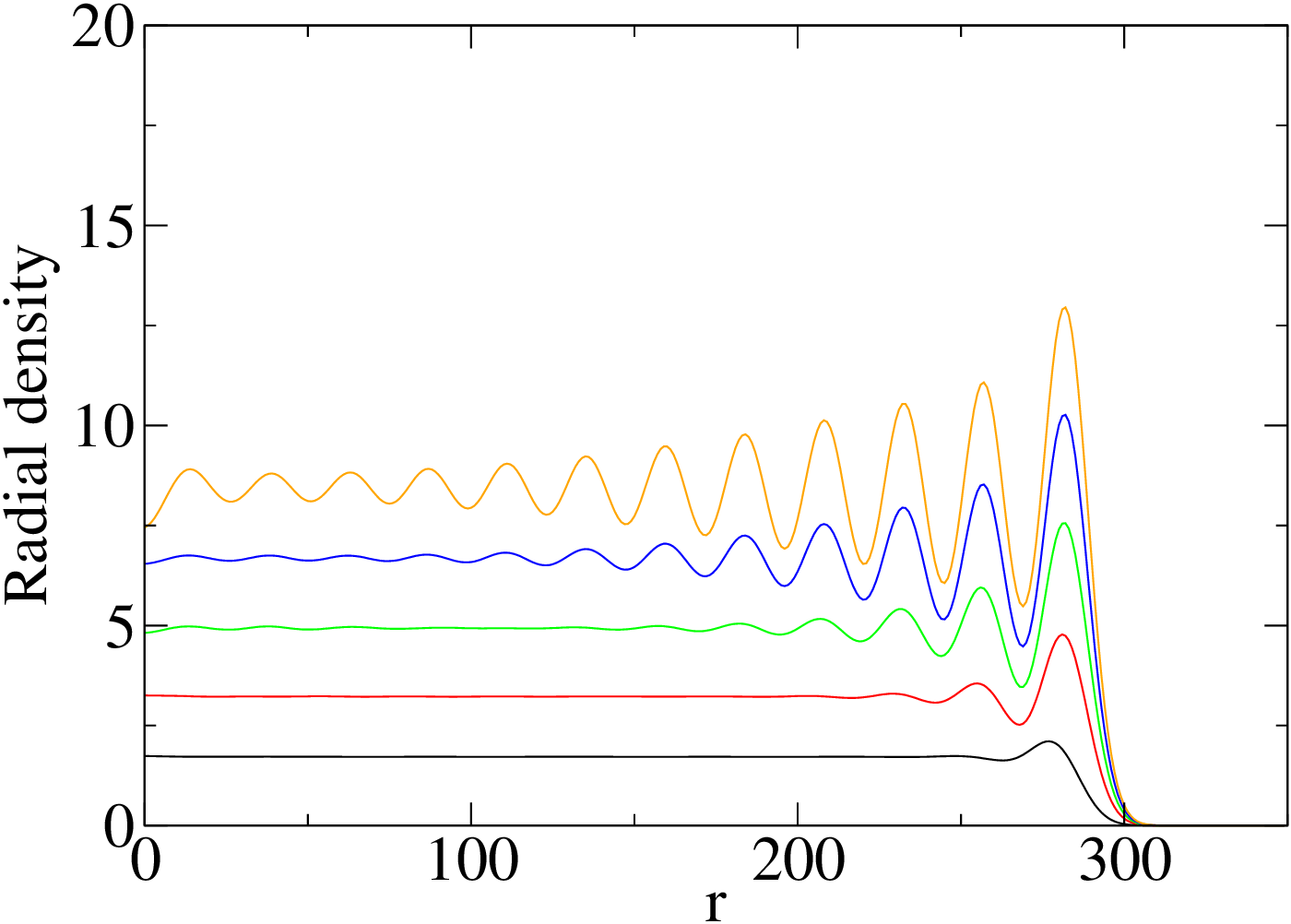}
\caption{Order parameter radial densities for $\alpha_{T}$ equal to
  $-2$, $-4$, $-6$, $-8$ and $-10$ from bottom to top. There are $400$ vortices in the disc.}
\label{fig:GLRD} 
\end{figure}  


We now consider the thermal averaged superconducting order parameter density $\langle|\psi
(r)|^{2}\rangle$ as a function of the distance from the center of the
sample $r$ in the GL LLL model. This can be computed from the
$N+1$ averages $\langle v_{m}v_{m}^{\star}\rangle$ as follows
\[ \sum_{n=0}^{N} \langle v_{n}v_{n}^{\star} \rangle
  \frac{r^{2n}}{n!2^{n}}   e^{-r^{2}/2}.   \]  When   $\alpha_{T}$  is
positive, the  order parameter density  is uniform except for  a sharp
drop to zero  at the edges of the sample. Plots  of the radial density
for various negative  values of $\alpha_{T}$ and $N=400$  are shown in
Fig. \ref{fig:GLRD}.   As $\alpha_{T}$ is  lowered, peaks form  at the
edges and increase in amplitude; shell-like structure spreads into the
sample from the  edges. At about $\alpha_{T}=-8$, we  observe that the
ripples have spread  all the way across the sample. If  we look at the
particle  configurations for  negative $\alpha_{T}$,  we see  that the
peaks are  due to increasingly shell-like ordering  towards the edges.
We have confirmed (for small system sizes for which we can compute the
vortex positions quickly) that the troughs in the radial density match
up  with  regions of  high  vortex density  as  we  would expect.  The
tendency  to form  a triangular  crystal at  low  temperatures emerges
quite  close to  the edge  of the  sample so  that the  evolution from
shell-like  order to  triangular order  happens within  a  few lattice
spacings. There is no evidence for  a phase transition to a crystal in
this system; instead the ordering is gradual \cite{ONeill}. The period
of the  radial density ripples  is the same  for all system  sizes and
temperatures. In fact, if we  compare the radial densities for samples
of different $N$ at a given temperature, we find that they very nearly
overlap close  to the edges for  $N=400$ and $N=600$,  but the $N=100$
discs show pronounced differences compared  to the other two sizes. We
have  repeated the  scaling analysis  that  we applied  to the  radial
density  peaks of  the  2D OCP,  this  time with  the assumption  that
$\xi\sim(|\alpha_{T}| -\beta)$ as suggested for $\alpha_{T}\leq -8$ in
work  by   Dodgson  and  Moore  \cite{Dodgson}.    For  less  negative
$\alpha_{T}$ we  use results for the correlation  length directly from
that paper. The fit is similar in appearance to Fig. \ref{fig:scaling}
but  we  do  not  reproduce  it  here; it  is  slightly  less
convincing than the scaling in the 2D OCP plasma.

We have  found for  the 2D OCP  in the  disc geometry that  the radial
density oscillations  allow us to  see directly the magnitude  of the
correlation length  in the bulk fluid.  The same behavior  is found in
the GL  LLL model.  It is  now clear that  those simulation studies which
reported  a  first-order  fluid-crystal  transition \cite{Leeuw,Choquard,Xing,Kato,Hu}   involved  systems   whose  linear
dimensions at  the reported  transition were actually  comparable with
the correlation length.  It would clearly be desirable to repeat these
studies for larger systems  before accepting the conclusion that 
there is a first-order transition in  these two-dimensional systems.

Finally, we note that it should be possible to see the density oscillations 
we have reported in this paper  at the edges of thin film type II superconducting samples.

PAM would like to acknowledge financial support from the EPSRC.


\begin{thebibliography}{99}

\bibitem{Leeuw}
J. W. Perram and S. W. de Leeuw, Physica {\bf 113A}, 546 (1982).

\bibitem{Choquard}
P. Choquard and J. Clerouin, Phys. Rev. Lett. {\bf 50}, 2086 (1983).

\bibitem{Moore}
M. A. Moore and A. P\'{e}rez-Garrido, Phys. Rev. Lett. {\bf 82}, 4078 (1999).

\bibitem{Lai}
Ying-Ju Lai and Lin I, Phys. Rev. E. {\bf 64}, 015601(R) (2001).

\bibitem{Xing}
L. Xing and Z. Te\v{s}anovi\'{c}, Physica  {\bf C196}, 241 (1992).

\bibitem{Kato}
Y. Kato and N. Nagaosa, Phys. Rev. B {\bf 47}, 2932 (1993).

\bibitem{Hu}
Jun Hu and A. H. MacDonald, Phys. Rev. Lett. {\bf 71}, 432 (1993).

\bibitem{Dodgson}
M.J.W. Dodgson and M. A. Moore, Phys. Rev. B {\bf 55}, 3816 (1997).

\bibitem{ONeill}
J.A. O'Neill and M. A. Moore, Phys. Rev. B {\bf 48}, 374 (1992).

\bibitem{Evans} R. Evans, R. J. F. Leote de Carvalho, J. R. Henderson and D. C. Hoyle, J. Chem. Phys. {\bf 100}, 591 (1994).

\bibitem{Evans2} R. Evans and R. J. F. Leote de Carvalho, J. Chem. Phys. {\bf 83}, 619 (1994).

\bibitem{Hansen} J. P. Hansen and I. R. McDonald, {\em Theory of simple liquids 2nd ed.}, Academic Press (1986).

\bibitem{Ruggeri}
G. J. Ruggeri and D. J. Thouless, J. Phys. F {\bf 6}, 2063 (1976).

\bibitem{Eilenberger}
G. Eilenberger, Physical Review {\bf 164},628 (1967).

\bibitem{Urbach}
J. S. Urbach, W. R. White, M. R. Beasley and A. Kapitulnik, Phys. Rev. Lett. 
 {\bf 69}, 2407 (1992).

\bibitem{Metropolis}
N. Metropolis, A. W. Rosenbluth and M. N.  Rosenbluth, A. H. Teller
 and E. Teller, J. Chem. Phys.  {\bf 21}, 1087 (1953).

\end{thebibliography}
\end{document}